\documentclass[article]{aipproc}
\layoutstyle{8x11double}
\usepackage{times}

\def\micron{$\mu$m}
\def\microG{$\mu$G}

\def\deg{$^\circ$}

\def\HI{H$^{\rm o}$}

\def\HeI{He$^{\rm o}$}

\def\nHI{$n(\mathrm{H}^0)$}

\def\npro{$n(\mathrm{p}^+)$}

\def\P5{$P_5$}

\def\kms{\hbox{km s$^{-1}$}}

\def\HeI{He$^{\rm o}$}

\def\cmtwo{cm$^{-2}$}

\def\cc{cm$^{-3}$}

\newcommand\apj{{ApJ}}% % Astrophysical Journal
\newcommand\apjl{{ApJ}}% % Astrophysical Journal, Letters
\newcommand\apjs{{ApJS}}% % Astrophysical Journal, Supplement
\newcommand\aap{{A\&A}}% % Astronomy and Astrophysics
\newcommand\nat{{Nature}}% % Nature
\newcommand\jgr{{J.~Geophys.~Res.}}% % Journal of Geophysical Research
\usepackage{graphicx}
\begin{document}   
%===============================================================================
%===============================================================================
\title{Exclusion of Tiny Interstellar Dust Grains From the Heliosphere}
\classification{96.50.Xy,95.30.Wi,96.50.Dj,96.50.Ek,98.38.Cp}
\keywords{heliosphere; interstellar matter}
%===============================================================================
\author{J.~D.~Slavin}{address={Harvard-Smithsonian Center for Astrophysics, 60
Garden St., Cambridge, MA  02138 (USA)}}
\author{P.~C.~Frisch}{address={Dept. Astronomy \& Astrophysics, U. Chicago,
5640 S.  Ellis Ave., Chicago, IL  60637 (USA)}}
\author{J.~Heerikhuisen}{address={U. Alabama-Huntsville, Center for Space
Plasma and Aeronomic Research, Huntsville, AL 35899}}
\author{N.~V.~Pogorelov}{address={U. Alabama-Huntsville, Center for Space
Plasma and Aeronomic Research, Huntsville, AL 35899}}
\author{H.-R.~Mueller}{address={Dept. of Physics and Astronomy, 6127 Wilder
Lab , Dartmouth College,Hanover, NH 03755}}
\author{W.~T.~Reach}{address={Infrared Processing and Analysis Center,
Caltech, Pasadena, CA 91125}}
\author{G.~P.~Zank}{address={U. Alabama-Huntsville, Center for Space Plasma
and Aeronomic Research, Huntsville, AL 35899}}
\author{B.~Dasgupta}{address={U. Alabama-Huntsville, Center for Space Plasma
and Aeronomic Research, Huntsville, AL 35899}}
\author{K.~Avinash}{address={U. Alabama-Huntsville, Center for Space Plasma
and Aeronomic Research, Huntsville, AL 35899}}

\begin{abstract}
The distribution of interstellar dust grains (ISDG) observed in the Solar
System depends on the nature of the interstellar medium-solar wind
interaction. The charge of the grains couples them to the interstellar
magnetic field (ISMF) resulting in some fraction of grains being excluded from
the heliosphere while grains on the larger end of the size distribution, with
gyroradii comparable to the size of the heliosphere, penetrate the termination
shock.  This results in a skewing the size distribution detected in the Solar
System. 

We present new calculations of grain trajectories and the resultant grain
density distribution for small ISDGs propagating through the heliosphere.  We
make use of detailed heliosphere model results, using three-dimensional (3-D)
magnetohydrodynamic/kinetic models designed to match data on the shape of the
termination shock and the relative deflection of interstellar \HI\ and \HeI\
flowing into the heliosphere.  We find that the necessary inclination of the
ISMF relative to the inflow direction results in an asymmetry in the
distribution of the larger grains (0.1 \micron) that penetrate the heliopause.
Smaller grains (0.01 \micron) are completely excluded from the Solar System at
the heliopause.
\end{abstract}

\maketitle

\section{Introduction}
Up to one percent of the mass of the interstellar cloud surrounding the
heliosphere is carried by dust grains that interact with the heliosphere
\citep{SlavinFrisch:2008,Landgraf:2000}.  The flow of interstellar material
past the Sun at 26.4 \kms\ drives large interstellar dust grains into the
heliosphere, while small grains are diverted around the heliosphere by the
interstellar magnetic field (ISMF).  The heliospheric trajectories of
intermediate sized grains can be complicated and depend on the solar wind
magnetic field and thus solar cycle phase.  Observations of ISDGs in the solar
system by the Ulysses, Galileo and Cassini spacecraft show that the density of
smallest grains are deficient in the inner heliosphere when compared to the
nominal Mathis et al. ``MRN'' power law size distribution
\citep[e.g.][]{Frischetal:1999,MRN:1977}.  For instance, the density of grains
of mass $10^{-14}$ g ($a \sim 0.1$ \micron) is reduced in the inner
heliosphere by a factor of $\sim 90$ below the MRN predictions and $10^{-15}$
g ($a \sim 0.05$ \micron) grains are deficient by three orders of magnitude.
At the same time large ($a \sim 1$ \micron) grains are absent from the MRN
distribution, but are abundant in the inflowing ISDGs.  While several
alternative grain size distribution models exist
\citep[e.g.,][]{Kim_etal_1994,Weingartner+Draine_2001,Zubko_etal_2004}, the
constraints of staying within the limits of cosmic abundances and explaining
the interstellar extinction curve lead in all cases to the presence in the
models of small grains that are absent from the observed distribution and a
cutoff on the large grain size end that is below that of the observed
distribution.  The deficiencies of small grains may result from small grains
being deflected around the heliosphere because their high charge-to-mass
ratios cause them to couple tightly to the ISMF. On the other hand, the
circumheliospheric interstellar medium (CHISM) may simply be deficient in
small grains (balancing the overabundance of large grains).  To evaluate which
is the case requires calculations of the grain distribution as a function of
grain size in the context of realistic heliosphere models. With recent data on
the shape of the heliosphere from the crossing of the termination shock (TS)
by Voyagers 1 and 2 \citep{Stone_etal_2008}, and on the deflection of
inflowing interstellar H$^0$ relative to He$^0$ \citep{Lallement_etal_2005},
such models face tighter constraints than ever before.  The inferred asymmetry
of the heliosphere can be explained by the orientation of the ISMF relative to
the direction of the cloud-Sun relative motion.  The ISMF required to fit the
heliosphere data, in turn, has direct implications for ISDG trajectories.

In this paper we present new calculations of grain trajectories and density
distributions for small grains that are primarily excluded from the inner
heliosphere.  Small grains experience enhanced charging rates due to the
ejection of secondary electrons in the hot $10^5 - 10^6$ K plasma between the
termination shock and heliopause \citep{KimuraMann:1998}.  We calculate grain
trajectories in 3-D, based on a self-consistent steady state MHD-kinetic
heliosphere model with grain charge calculated at each location along the path
according to the plasma conditions and radiation field. (We note that
the recent Voyager 2 results \citep{Richardson_etal_2008} on the lower than
expected plasma temperature in the inner heliosheath are not incorporated in
the heliosphere model we use.  We expect that this could result in somewhat
lower dust charging in this region than we calculated, though the detailed
effects of charging by thermal particles vs.\ PUIs has yet to be fully
explored.)  The model of plasma density and magnetic field in the heliosphere
at each point of space properly accounts for the asymmetries that result from
the angle between the ISMF and interstellar gas/dust inflow direction, and
ISMF and ecliptic plane \citep{PogorelovStoneetal:2007,Pogorelov_etal_2008}.
When combined with recent models of grain charging
\citep{WeingartnerBarr:2006} and a realistic ultraviolet (UV) radiation field,
the Lorentz force at each point in space can be used to calculate the grain
trajectory.  Total 3-D grain densities are then calculated for a sample of
interstellar grain sizes based on the total gas-to-dust mass ratio in the
CHISM and ISDG models.  Many previous calculations have been made of grain
interactions with the heliosphere
\citep[e.g.][]{Grun_etal_1994,Frischetal:1999,KimuraMann:1998, Landgraf:2000}
but ours is the first to use a realistic heliosphere model, including
distortion due to an ISMF orientation that is consistent with recent
observational data.

\section{Grain Trajectory Calculations}
\subsection{Heliosphere Model}
\begin{figure}
\includegraphics[width=0.45\textwidth]{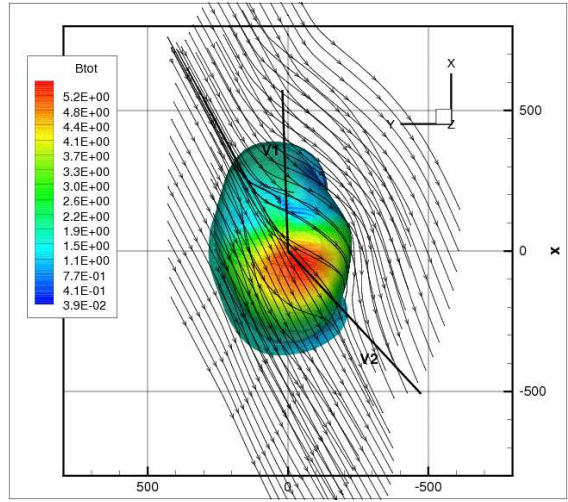}
\caption {3-D MHD/kinetic heliosphere model used in these calculations
\citep{PogorelovStoneetal:2007,Pogorelov_etal_2008}, showing the heliopause
boundary, magnetic field lines and the Voyager 1 and 2 trajectories. The
viewpoint is from interstellar space looking towards the downwind. The
heliopause is colored according to the strength of the magnetic field. The
bending of the magnetic field lines around the heliopause
dominates the trajectories of tiny dust grains interacting with the
heliosphere, and is responsible for the dust trajectories and hence the
enhanced dust densities.}
\label{fig:np}
\end{figure}

The grain trajectory calculations require a 3-D model that self-consistently
includes the interaction and charge exchange between interstellar H and the
solar wind plasma, and that is based on realistic boundary conditions for the
heliosphere as set by the surrounding interstellar cloud.  The 10 AU
difference in the termination shock distances seen by Voyager 1 versus Voyager
2 suggests an asymmetry in the outer heliosphere topology that can be
explained by the tilt of the ISMF with respect to the upwind direction.  We
therefore select a heliosphere model that reproduces the TS asymmetry, the
hydrogen-helium offset of $\sim 5$\deg, and the angle of the ISMF with respect
to the ecliptic plane as indicated by the H-He offset and data on the ISMF
near the Sun.  We use the heliosphere of
\citet{PogorelovStoneetal:2007,Pogorelov_etal_2008} and
\citet{Heerikhuisen_etal_2008}, which is based on coupled 3-D MHD (ion) fluid
and kinetic (neutral) particle code including the effects of charge exchange
to calculate the plasma and neutral particle properties.  The model parameters
of the solar wind and inflowing interstellar material are given in Table 1
(where \nHI\ is the neutral H density and \npro\ is the proton density).  The
solar wind in the model carries a Parker spiral type magnetic field and has
$v(\mathrm{H}^+)_\mathrm{SW} = 450$ \kms.
%: $V_\mathrm{ISM} = 26.4$ \kms, $n(\mathrm{H}^+)_\mathrm{ISM} = 0.06$ \cc,
%$T_\mathrm{ISM} = 6500$ K,
%$n(\mathrm{H}^0)_\mathrm{ISM} = 0.15$ \cc, 
The ISMF strength is $B_\mathrm{ISM} = 3$ \microG\ with the direction 30\deg\
away from the ISM inflow direction and pointed toward the southern hemisphere,
and lying in the H deflection plane for this model \citep{Pogorelov_etal_2008}.
The heliosphere model uses a coordinate system with the x-axis perpendicular
to the ecliptic plane, the z-axis in the ecliptic plane towards upwind and
perpendicular to the x-axis, and the y-axis in the ecliptic plane to form a
righthanded coordinate system. The Sun is located at $x = y = z = 0$.  Fig.
\ref{fig:np} shows the ISMF lines bending over the heliopause.

\begin{table}
\begin{tabular}{lccccc}
\hline
 &{Velocity} &{$n(\mathrm{p^+})$} &{$n(\mathrm{H^\circ})$}   &{Temperature}   &{Magnetic Field}   \\
 &{\kms} &{\cc} & {\cc}   &{K}   &{\microG}   \\
\hline
ISM at ``$\infty$'' & 26.4 & 0.06 & 0.15  & 6,527 & 3.0 \\
Solar Wind at 10 AU & 454 &  0.07  & 0  & $2,750-14,280$ & $0.4-3.3$ \\
\hline
\end{tabular}
\caption{Parameters of Heliosphere Model with Positive Magnetic Polarity at
the  NEP}
\label{tab:model}
\end{table}

\begin{figure*}[ht!]
\includegraphics[width=0.45\textwidth]{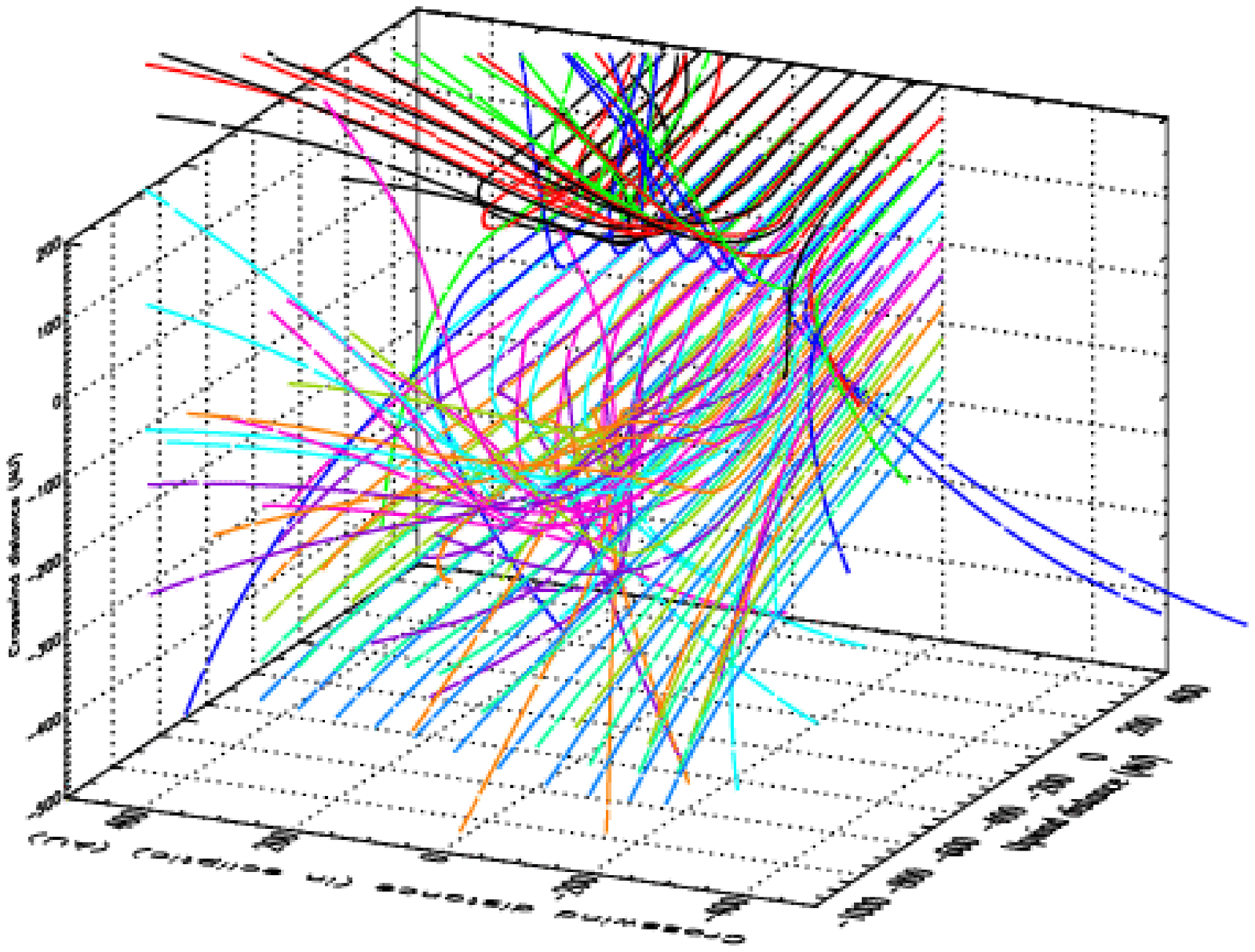}
\includegraphics[width=0.45\textwidth]{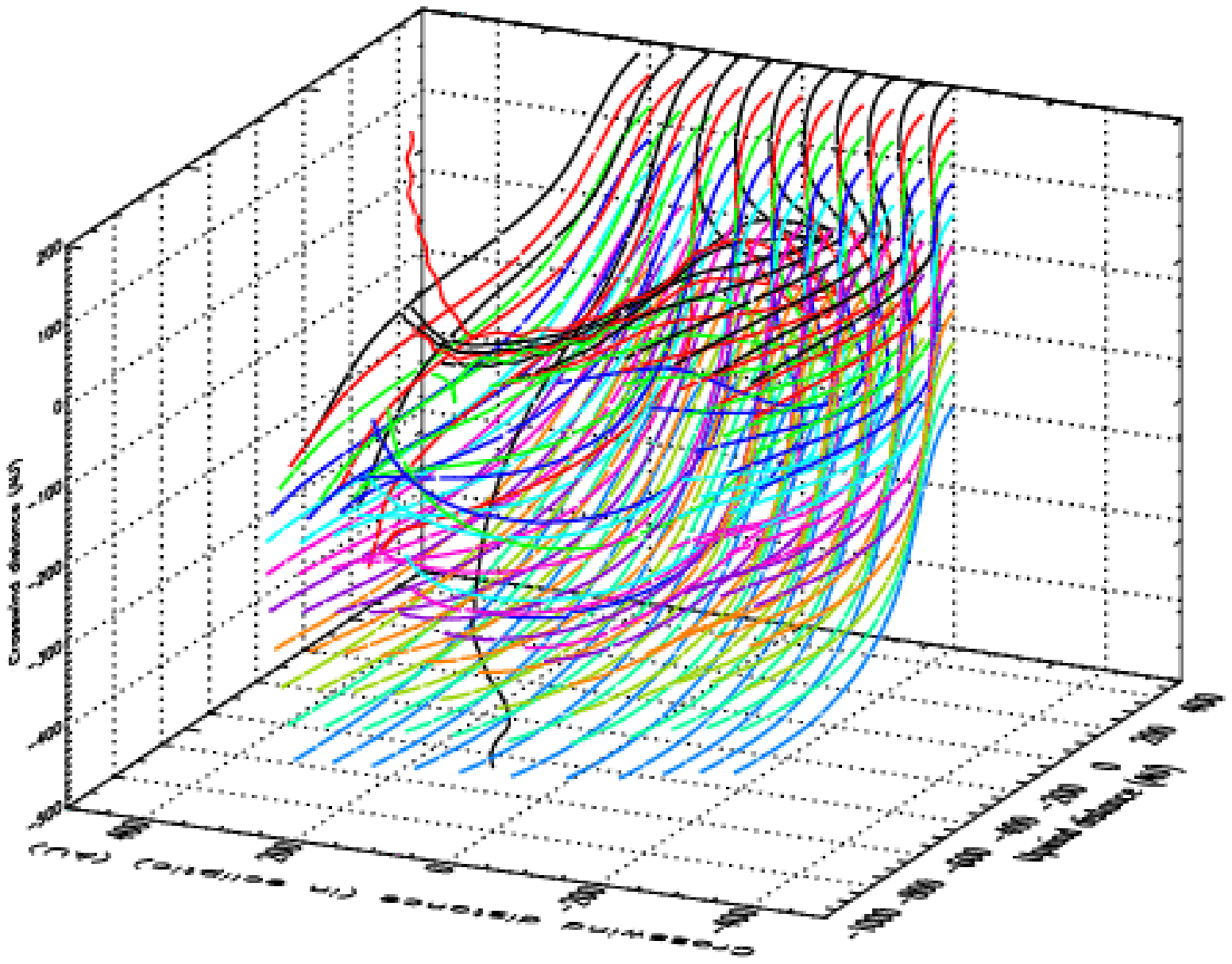}
\caption{
Example of spatial trajectories for 0.1 \micron\ grains (left) and
0.01 \micron\ grains (right). The upwind direction is toward the back plane
while the ecliptic is parallel to the bottom plane. Note the convergence of
trajectories in the top and bottom, with most 0.01 \micron\ grains deflected
to low latitudes.  The 0.01 \micron\ grains are entirely deflected around the
heliosphere.}
\label{fig:traj}
\end{figure*}

\subsection{Grains and Grain Charging}
We illustrate the deflection of small grains around the heliosphere by
calculating the charge and trajectories for spherical grains of radii 0.1
\micron, and 0.01 \micron, composed of astronomical silicates.  The assumption
that the small grains are silicates is supported by observations and analysis
indicating that the gas phase carbon abundance in the CHISM is high
\citep[supersolar in fact][]{Slavin+Frisch_2006} leaving none to be locked up
in grains.  Grain charging is calculated using the model and grain-charging
code of \citet{WeingartnerBarr:2006}, combined with plasma densities and
temperatures from the heliosphere model, and the modeled extreme-UV/far-UV
photon fluxes as discussed below.

\begin{figure*}[ht!]
\includegraphics[width=0.45\textwidth,clip=true]{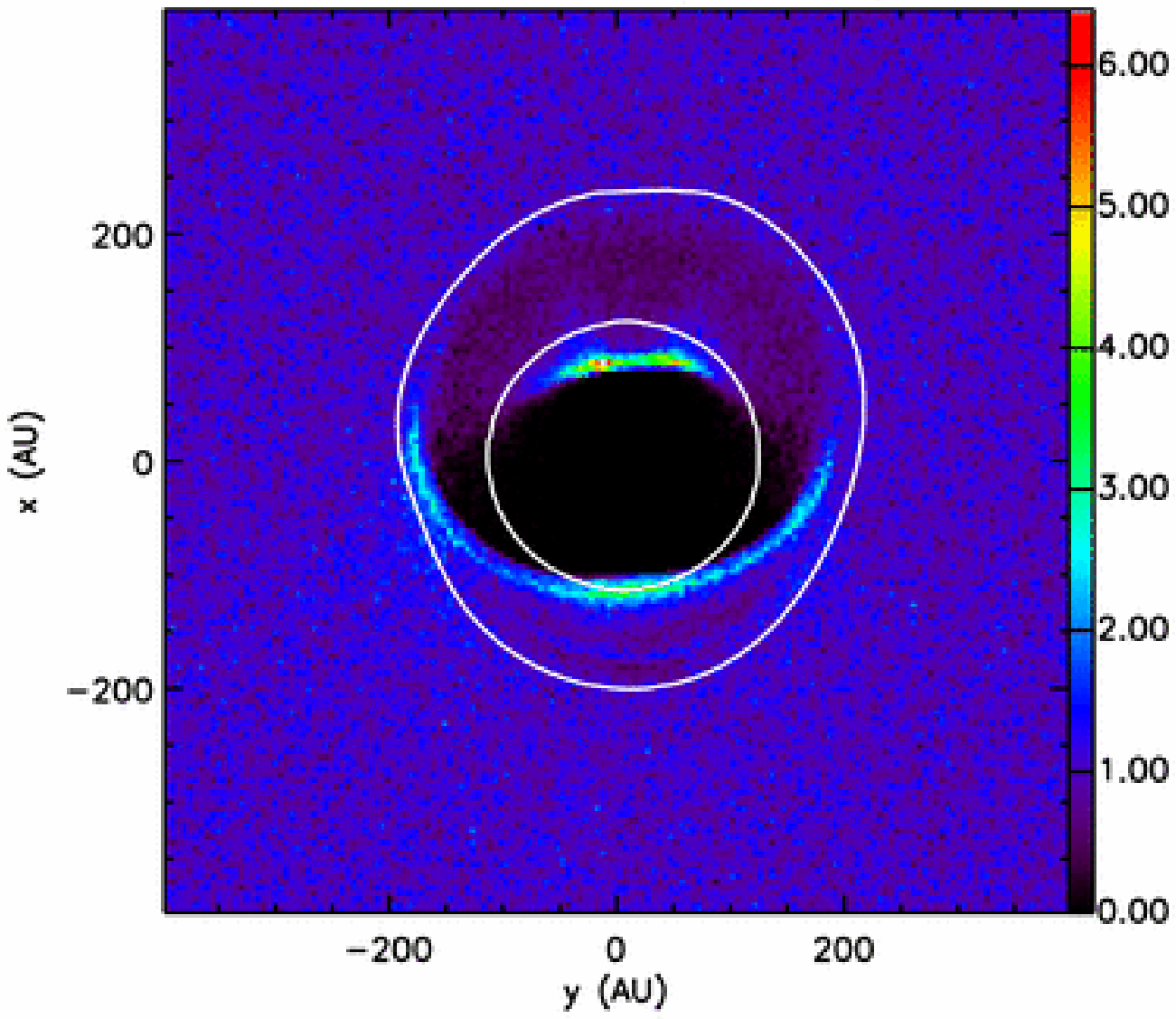}
\includegraphics[width=0.45\textwidth,clip=true]{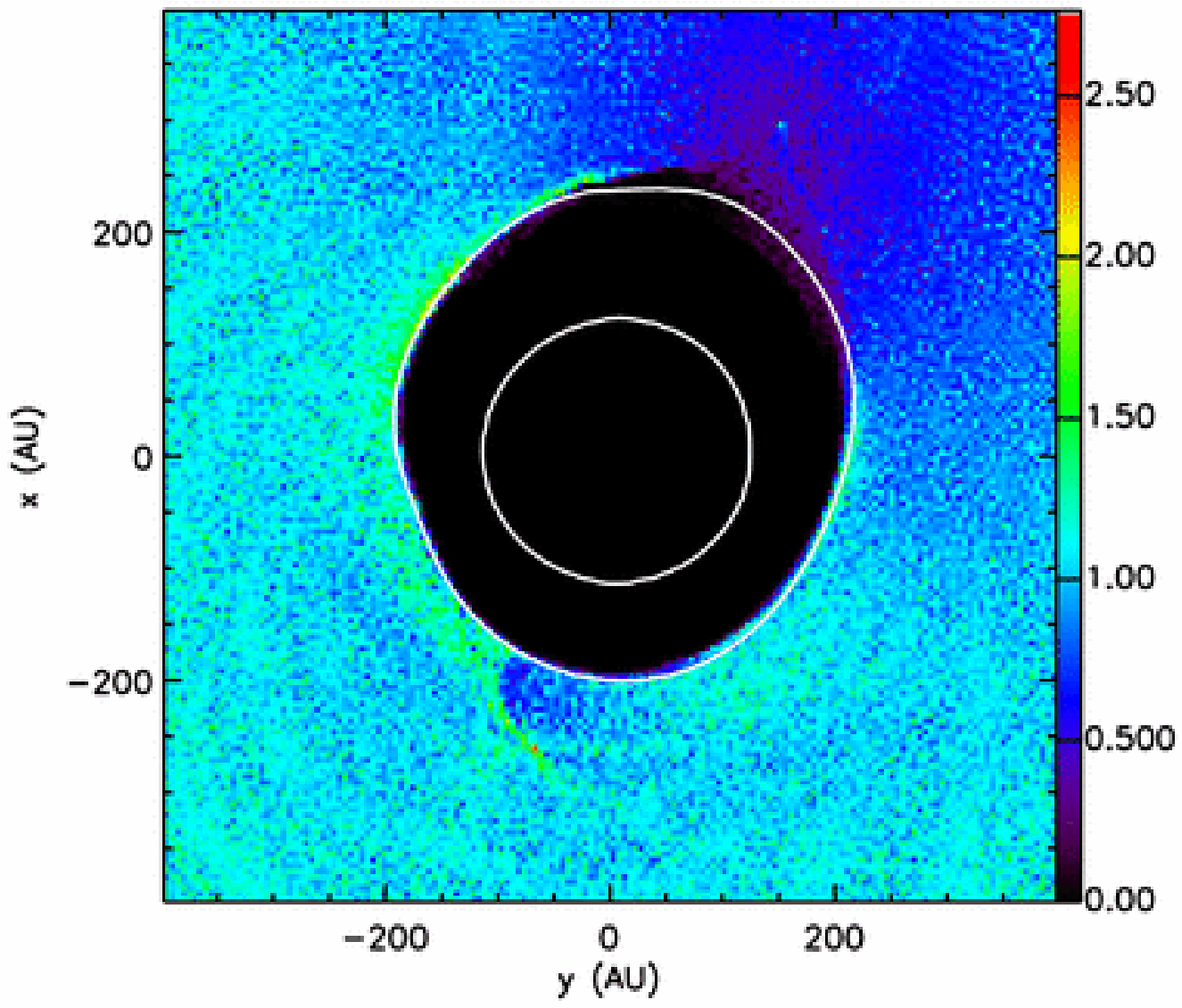}
\caption{Dust density distributions (\cc) in a slice in the z = 0
plane (perpendicular to the IS flow direction) for radius 0.1 \micron\
(left figure) and 0.01 \micron\ (right figure) silicate grains. (See
text for coordinate system.)  These views are centered near the upwind
direction, and ecliptic longitude increases towards the left. The
inner nearly circular contour is the solar wind termination shock,
while the outer contour is the heliopause.  The $a \sim 0.1$ \micron\
grains are partially diverted at the heliopause, but some grains
penetrate inside the TS. All 0.01 \micron\ grains, which have small
gyroradii due to a high $q/m$ ratio, are excluded from the 
heliosphere.  The colors indicate the dust
density relative to its value in the pristine ISM. Note that the
plotted density range is much larger for the 0.1 \micron\ grains.}
\label{fig:zplane}
\end{figure*}

\begin{figure*}[ht!]
\includegraphics[width=0.46\textwidth,clip=true]{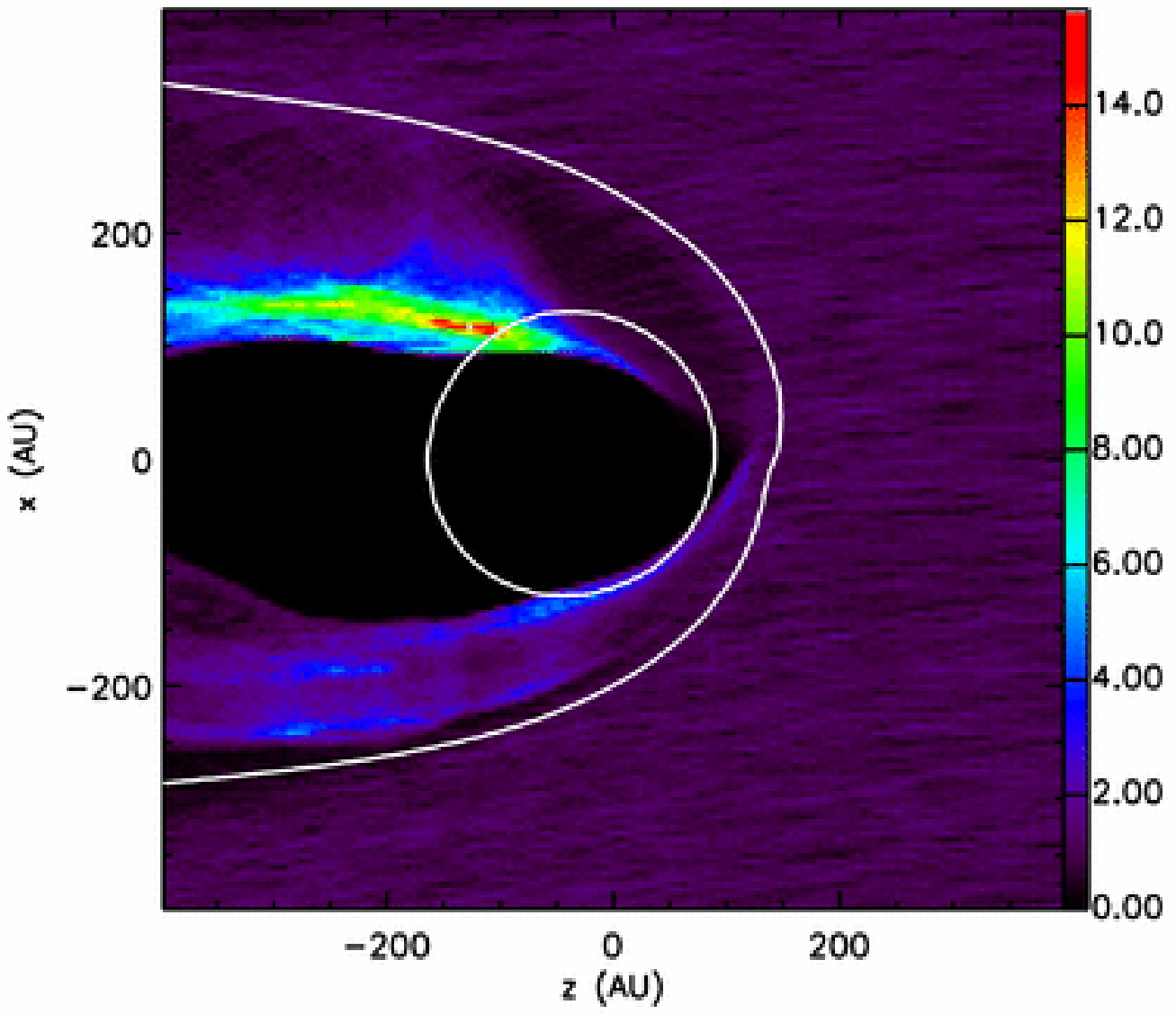}
\includegraphics[width=0.46\textwidth,clip=true]{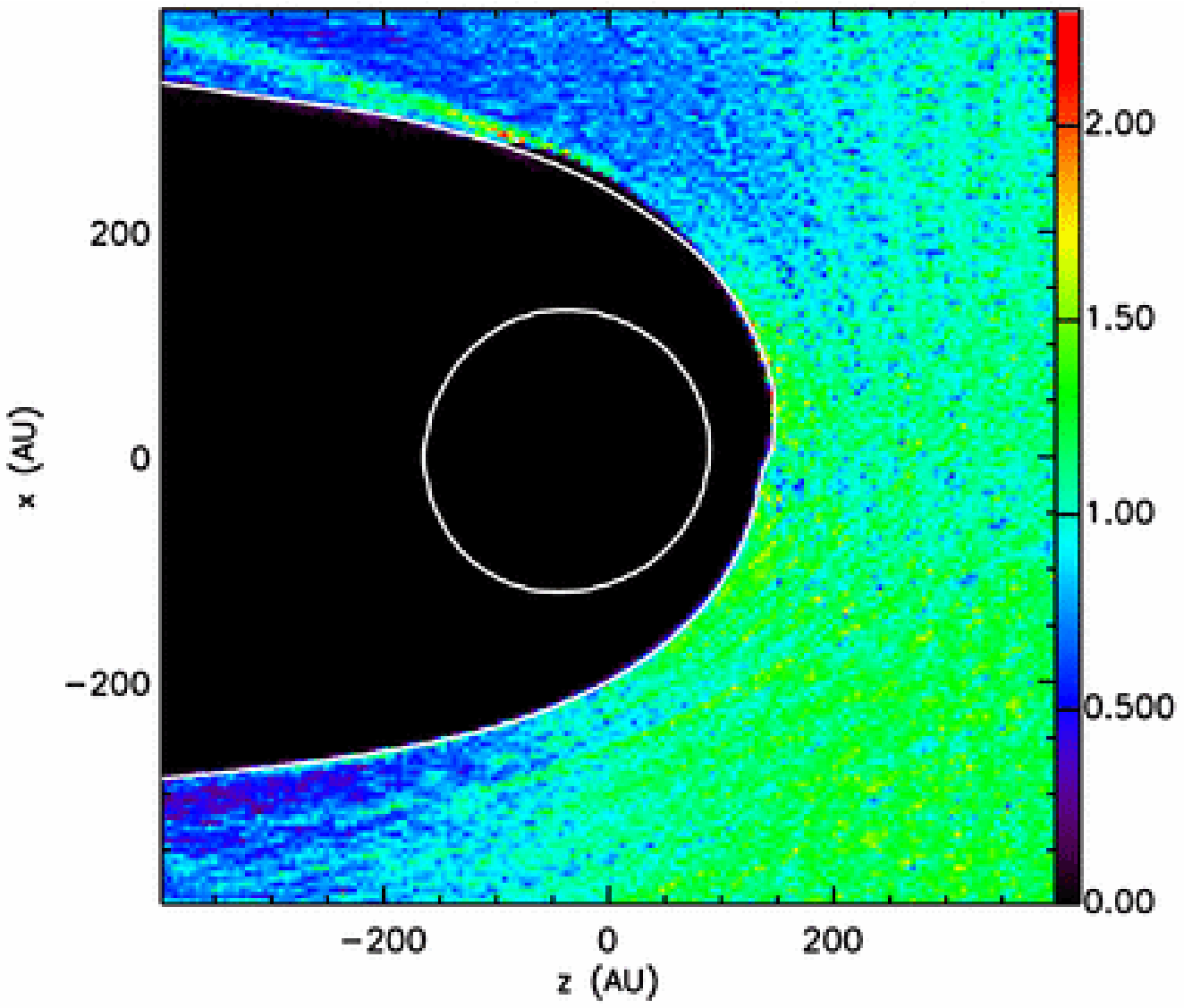}
\caption {Spatial densities of grains (\cc) for slices in the xz
plane. The right figure (0.01 \micron\ grains) shows the full
exclusion of tiny grains from the heliosphere. The left figure (0.1
\micron\ grains) shows the diversion of slightly larger grains able to
partially penetrate the inner heliosheath. The concentration of grains
near the North Ecliptic Pole results from the deflection of particles
in the inner heliosheath where they are sensitive to the magnetic
polarity of the solar wind.}
\label{fig:xyplane}
\end{figure*}

The primary mechanisms affecting the charge of grains near and in the
heliosphere are the photoelectric emission caused by absorption of photons
from the UV radiation field, and the sticking of electrons and ions that
collide with the grains.  Additional grain charging occurs in the hot subsonic
solar wind plasma in the inner heliosheath between the solar wind termination
shock and the heliopause because of secondary electron emission produced by
collisions of high energy electrons with the grains.

The radiation field used in our calculations is comprised of the constant
interstellar radiation field from \citet{Gondhalekar_etal_1980}, and the Solar
flux data taken from the TIMED/SEE project (see
\url{http://lasp.colorado.edu/see/see_data.html}).  The TIMED/SEE project
creates spectra using a combination of measured and modeled fluxes. 
% In Fig.  \ref{fig:radfld}  we illustrate this radiation field.
The photon flux longward of 13.6 eV is dominated by the interstellar radiation
field, while the extreme ultraviolet radiation field is dominated by the Sun.
The interstellar and solar fluxes have equal effect in charging grains at
$\approx 100$ AU.

\subsection{Grain Trajectories}
The grain trajectory calculations begin with the grains upstream of the
heliosphere and traveling with the interstellar gas. However, the grains are
likely to have been accelerated by turbulence in the interstellar magnetic
field and thus should have some initial velocity relative to the gas. Taking
guidance from \citet{Yan_etal_2004}, we give the grains an initial 3 \kms\
speed (roughly the gas turbulent velocity observed for the Local Interstellar
Cloud) relative to the gas, and with a net direction that is perpendicular to
the magnetic field but is otherwise randomly directed.

We integrate the equations of motion for the grains, calculating at each point
the charge on the grain depending on its position (and properties of the
surrounding plasma) and including the Lorentz forces, radiation pressure and
gravity of the Sun.  For the density distribution we take over $10^6$ grain
trajectories for grains starting in a grid in a plane far ($\sim 900$ AU) from
the Sun in the ISM.

The forces acting on the grains include gravity, radiation pressure, and the
Lorentz force from the motion of charged grains relative to the magnetic field
lines.  The equation of motion is given by 
\begin{equation}
\ddot{\vec{x}}_\mathrm{g} + (1 - \beta) \frac{G
M_\mathrm{sun}}{|\vec{x}_\mathrm{g}|^3} \vec{x}_\mathrm{g}
- \frac{q}{m} (( \dot{\vec{x}}_\mathrm{g} -  \vec{v}_\mathrm{gas}) \times
\vec{B}) = 0
\end{equation}
where $\dot{\vec{x}}_\mathrm{g}$, $q$, and $m$ are the velocity, charge and
mass of the grain, $\beta$ is the ratio of radiation pressure to gravitational
attraction, $\vec{B}$ is magnetic field, and $\vec{v}_\mathrm{gas}$ is gas
(solar wind or ISM) velocity.  Although radiation pressure is not significant
for spherical grains in the outer heliosheath that obey Mie scattering,
\citet{KimuraMann:1998fluffy} have shown that radiation pressure acting on
fluffy grains results in non-radial torques that may significantly perturb the
grain motion in the outer heliosphere.  This effect is not included here.

We show sample grain trajectories for silicate grains of radii 0.01 \micron\
and 0.1 \micron\ in Fig. \ref{fig:traj}.  The resulting density distributions
of grains (\cc) are shown for a slice in the plane at $z=0$ that is
perpendicular to the interstellar flow direction (Fig. \ref{fig:zplane}), and
for a meridian slice through the heliosphere nose (Fig. \ref{fig:xyplane}).

These simulations of grain locations in the extended 3-D volume were performed
with the University of Alabama, Huntsville, CSPAR cluster.

\section{Results}
\begin{figure}[ht!]
\includegraphics[width=0.48\textwidth,clip=true]{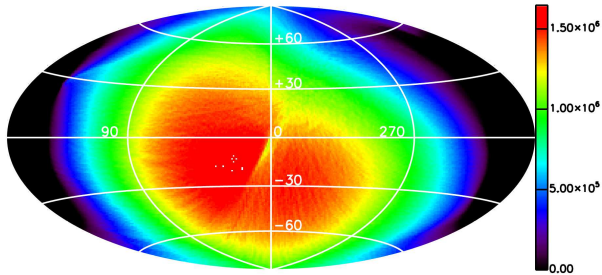}
\caption{Column densities (\cmtwo) of 0.01 \micron\ grains deflected
around the heliopause, out to a distance of 400 AU from the Sun.  The values
on the color bar show the true column density (\cmtwo) for the assumptions
that the gas-to-dust ratio is 100 and all the grains are 0.01
\micron.  The upwind direction is at the center, with 0\deg corresponding to
ecliptic coordinate $\ell \sim 255$\deg.  
}
\label{fig:coldens}
\end{figure}

This study shows that the distortion of the local interstellar
magnetic field by the heliosphere causes large-scale asymmetries in
the spatial distribution of tiny interstellar grains flowing past the
Sun and within 1000 AU.  Grains of radius $\sim 0.01$ \micron\, appear
to be completely excluded from the heliosphere.  The trajectories of
these grains will be sensitive to the direction of the local
interstellar magnetic field, and relatively insensitive to the solar
22-year magnetic activity cycle since the grains do not penetrate the
heliopause.  In Fig. \ref{fig:coldens} we show the column densities of
these grains within 400 AU of the Sun, and outside of the heliopause.
The ISMF distortion by the heliosphere is imprinted on the
distribution of these small grains.

Grains as small as 0.1 \micron\ can partially penetrate the termination shock.
The trajectories of 0.1 \micron\ grains are initially influenced by the
interstellar magnetic field.  Once inside of the heliosphere, the solar wind
magnetic field and solar cycle become significant
\citep[e.g.][]{Landgraf:2000}.  The high charge-to-mass ratio for these grains
is enhanced by secondary electron ejection in the inner heliosheath,
increasing the coupling between 0.1 \micron\ grains and the magnetic field
in this region.

\begin{theacknowledgments}
We thank Joe Weingartner for sharing his grain-charging code with us,
and Richard Lieu, Haru Washimi, and Xianzhi Ao for helpful comments.
This work has been supported by NASA grants NNX08AJ33G, NNX09AH50G,
NNX07AH18G, NNX08AJ21G, NNX09AB24G,
NNX09AG29G, and NNX09AG62G, and NASA contract NNG05EC85C.
\end{theacknowledgments} 

\bibliographystyle{aipproc}

\end{document}